\documentclass[preprint,12pt]{elsarticle}

\usepackage{setspace}
\usepackage{tocloft}
\usepackage{longtable}
\usepackage{lscape}
\usepackage{afterpage}
\usepackage{tabularx}
\usepackage[notlot,notlof,nottoc,notindex]{tocbibind}
\usepackage{titlesec}
\usepackage{booktabs}
\usepackage{footmisc}
\usepackage{multirow}
\usepackage{amsmath}
\usepackage{amsfonts}
\usepackage{amssymb}
\usepackage{float}
\usepackage{comment}
\usepackage{parskip}
\cftpagenumbersoff{table}
\usepackage{latexsym}
\usepackage{makecell}
\usepackage{array, caption, threeparttable}
\usepackage{amssymb}
\usepackage{amsmath}
\usepackage{upgreek}
\usepackage[colorlinks,
linkcolor=blue,
 anchorcolor=blue,
 citecolor=blue]{hyperref}
\usepackage{epstopdf}

\journal{Carbon}

\begin{document}
\begin{sloppypar}
\begin{frontmatter}





\title{High-sensitivity graphene MEMS force and acceleration sensor based on graphene-induced non-radiative transition}
\author[label1]{Guanghui  Li}
\author[label2,label3]{Fengman  Liu}

\author[label1]{Shengyi Yang}

\author[label1]{Jiang-Tao Liu}
\ead{jtliu@semi.ac.cn}
\author[label1]{Weimin Li}
\ead{657378330@qq.com}

\author[label2,label3]{Zhenhua Wu}
\ead{wuzhenhua@ime.ac.cn}
\address[label1]{School of Physics and Mechatronic Engineering, Guizhou Minzu University, Guiyang 550025, China}
\address[label2]{Key Laboratory of Microelectronic Devices and Integrated Technology, Institute of Microelectronics, Chinese Academy of Sciences, Beijing 100029, China}

\address[label3]{School of Integrated Circuits, University of CAS, Beijing 100049, China}

\begin{abstract}
The micro-electromechanical-system (MEMS) force and acceleration sensor utilizing the graphene-induced non-radiative transition was investigated.
The graphene-induced non-radiative transition is very sensitive to the distance, and the deflection of the graphene ribbon is highly susceptive to applied force or acceleration.
Thus, a high-sensitivity MEMS sensor can be achieved with detecting the graphene ribbon's deflection of 1 $nm$, the force of 0.1 $pN$, and the acceleration of 0.1 $mg$.
The MEMS sensor, with a size of only tens of microns, can be charged by light irradiation without connecting power sources.
In addition, it allows long-distance detection, i.e., wireless transmitter circuit can be omitted.
Therefore, it will have significant application prospects in the fields of micro-smart devices, wearable devices, biomedical systems, and so on.
\end{abstract}


\end{frontmatter}



\section{Introduction}

With the development of intellectualization and wearable devices, high-precision micro-sensors have become the focus of researchers' attention\cite{D2LCH,ASGHG43,lu20hi,deing}. Many sensors convert the measured signal into deformation, such as pressure sensors\cite{DEVI2D,caoarab,sharmane,chun2,davidoj}, acceleration sensors\cite{Fanphe,Fnded,kaelop,HAurne,LISCD1}, and so on. Thus, the measurement of small deformation becomes the focus of the design of sensors. Optical fiber sensors are susceptible to small deformation and have high precision\cite{TAVB2d,VIGHD1,ZRang}. But they also have disadvantages, such as large size and difficulty in integrating with wearable devices.

Recently, micro-electromechanical system (MEMS) sensors, especially graphene MEMS sensors, have attracted researchers' attention. Graphene possesses an extremely thin thickness\cite{Leeasu}, superior mechanical properties\cite{tsateriz}, and unique optical\cite{maknt} and electrical properties\cite{boligh}, which makes it an ideal material for MEMS sensors. Most traditional graphene MEMS sensors convert external excitations such as pressure to graphene strain, and then achieve the sensing function by measuring the change in resistance caused by graphene strain\cite{caoarab,chun2,Smit20p}. However, the resistance of graphene changes greatly only when the strain is larger\cite{furain,zhwow,renhkj}, which reduces sense sensitivity, and the larger strain tends to make the graphene fracture. Furthermore, as these sensors are implemented to measure electrical signal for non-contact sensing, the appropriate circuitry, batteries, and wireless transmitter are required to be integrated. So, it is difficult to further reduce the size of the sensors, and the battery needs to be replaced frequently.

In this paper, a MEMS force and acceleration sensor was designed based on the feature that graphene-induced non-radiative transition is very sensitive to distance\cite{gaursal,swance,swaonance}. The non-radiative transition induced by graphene means that when graphene is placed in the near field of the fluorescent emitter\cite{nikds,mazzle,brenast}, it is possible that the NVC electron transition will not radiate photons, but will stimulate the electron in graphene to jump up through non radiative energy transfer. The probability of non-radiative transition of the emitter will increase dramatically\cite{nenic,velipro,reanical}. The smaller the distance between the emitter and graphene, the greater the probability of non-radiative transition, and the emission intensity will be significantly reduced.

The results show that when the graphene ribbons are deflected to 1 nm in this sensor, the graphene-induced non-radiative transition will reduce the emission intensity by a maximum of 4.13\%, achieving 0.1 pN force (1\% of the gravity of one living mammalian cell\cite{martial,adeeous}), or 0.1 mg acceleration detection. In addition, because temperature, sound wave, current, mass, and other changes can also cause MEMS to have small deflections, the sensing and detection of these signals can also be realized using the high sensitivity of graphene-induced non-radiative transitions to distance. This kind of sensor, with the size of only tens of microns, need not connect the circuit or the power supply, and can be charged by light irradiation. Because fluorescence can be detected from a long distance, no additional antenna need to be integrated, which makes the MEMS sensor easy to incorporate into organisms such as the lower part of the skin. It will have broad application prospects in the fields of micro-intelligent devices, wearable devices, biomedical systems and so on.

\section{Models and Theories}


\begin{figure}[t]
\begin{center}
\includegraphics[width=0.85\textwidth]{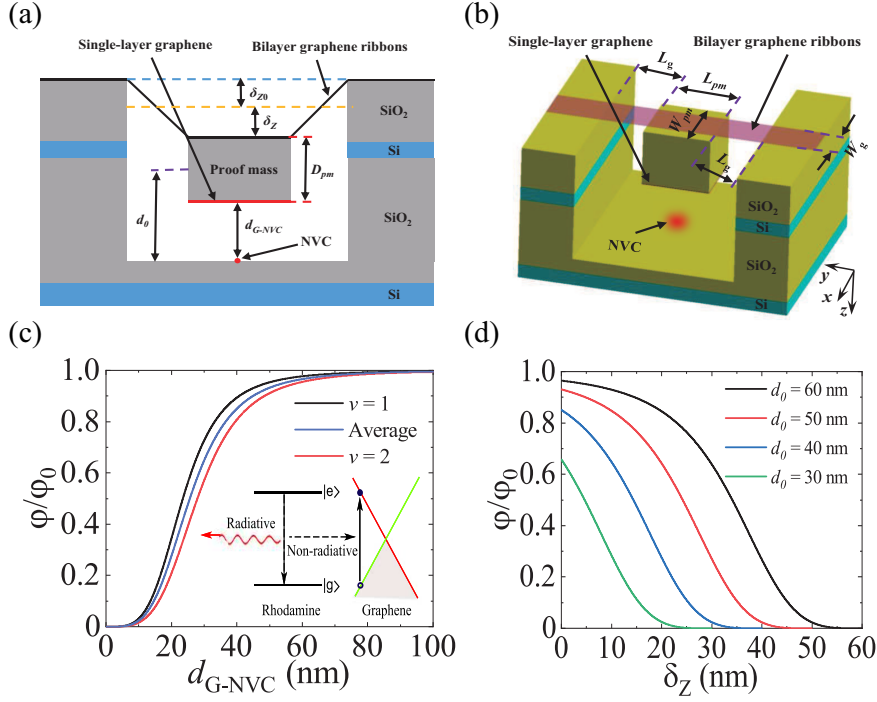}
\end{center}
\caption{Schematic of  the device structure (a) profile (b) overlooking. (c) Change of the emission intensity with the distance between the single-layer graphene and NVC in different emitter dipole directions, the inset shows the  schematic diagram of graphene-induced non-radiative transition. (d) Change of the emission intensity of NVC with the downward moving distance of proof mass at different initial distances between the proof mass and NVC.}
\label{fig1}
\end{figure}

The schematic diagram of the MEMS force sensors shown in Fig. \ref{fig1}(a), (b). The silica proof mass is suspended above the column through double-layer graphene ribbons. A single-layer graphene is attached to the lower surface of the proof mass. A fluorescent emitter, such as nitrogen-vacancy centers (NVC) in the diamond film, is placed under the proof mass. When a force is applied to the proof mass along the z-axis, the suspended double-layer graphene ribbons elongate, and the proof mass moves in the axial direction. Due to the near-field interaction between the single-layer graphene attached to the lower surface of the proof mass and NVC, it is possible that the electron transition of NVC will not release photons, but will make the electrons in graphene jump up through non radiative energy transfer. The excited electrons in graphene will return to the ground state through a non radiative process.  The probability of a non-radiative transition of NVC will be changed, resulting in a change in the emission intensity of NVC. Hence, by measuring the intensity of the NVC emission, the magnitude of the force can be obtained. Specifically, the change of emission intensity caused by graphene-induced non-radiative transition can be described as\cite{gaursal}:
\begin{equation}
{\varphi _G} = \frac{{{\varphi _0}}}{{1 + \frac{{9\nu \alpha }}{{256{\pi ^3}{{\left( {{\varepsilon _{er}} + 1} \right)}^2}}}{{\left( {\frac{\lambda}{{{d_{G - NVC}}}}} \right)}^4}}}
\end{equation}

Here, ${\varphi_{0}}$ is the emission of a single emitter in the absence of graphene,  $\nu  = 1$($\nu  = 2$) when the graphene surface is parallel(perpendicular) to the direction of the emitter dipole, ${\alpha}$ is the fine structure constant, ${\varepsilon _{er}}$ is the equivalent relative permittivity of the device, ${\lambda }$ is the emission wavelength of the emitter in free space, ${d_{G - NVC}} = d_0 - \delta_Z$ is the distance between the single-layer graphene and the emitter, ${d_0}$ is the initial distance between the single-layer graphene and the emitter, and ${\delta_Z}$ is the displacement of the proof mass under the action of force. The influence of different orientations of emitter dipoles on the emission intensity of NVC was studied.

When the emission wavelength of NVC is 638 nm, if the distance between the proof mass and NVC is close to 100 nm, the single-layer graphene has a little inhibition on NVC emission[Fig. \ref{fig1}(c)], and the emission intensity is close to that without graphene. Due to the large size of the proof mass, the change of the emission intensity of NVC is entirely caused by the single-layer graphene attached to the lower surface of the proof mass, and the effect of the suspended double-layer graphene ribbons is ignorable. When the graphene surface is parallel to the emitter dipole direction ($\nu = 1$), the emission intensity of NVC is larger than that of their vertical situation ($\nu = 2$) at the same distance. Since many NVC polarizations are randomly distributed, the total emission intensity is approximately replaced by the average of $\nu = 1$ and $\nu = 2$.

Since the emission intensity of NVC is strongly dependent on the distance ${d_{G - NVC}}$ between the single-layer graphene and NVC, the initial distance $d_0$ significantly affects the measurement range and sensitivity of the device. Fig. \ref{fig1}(d) shows the relationship between the emission intensity of NVC and the displacement of proof mass under different $d_0$ values. The larger $d_0$, the smaller the probability of graphene-induced non-radiative transition. The emission intensity of NVC changes dramatically with the displacement of proof mass. When the displacement of proof mass is 1 nm, the maximum emission intensity can reach 4.13\%.

\section{Results and discussion}
\subsection{Force sensor}
This structural mechanics model can be simplified as graphene ribbons fixed at both ends. In this way, the displacement of the proof mass caused by the force can be described by\cite{Fanphe}
\begin{equation}
{F + M_{pm}\rm g = 2\left( {\frac{{E_{\rm g}W_{\rm g}D_{\rm g}^3}}{{L_{\rm g}^3}}} \right)Z + \left( {\frac{{E_{\rm g}W_{\rm g}D_{\rm g}}}{{L_{\rm g}^3}}} \right){Z^3} + 2\left( {\frac{{{\sigma _0}W_{\rm g}D_{\rm g}}}{{L_{\rm g}}}}\right)Z}
\end{equation}

where, ${F}$ is the force applied externally, ${M_{pm}}$ is the mass of the proof mass, $\rm g$ = 9.80665 ${\rm m/{s^2}}$ is the gravitational acceleration, ${E_{\rm g}}$ is the Young 's modulus of the graphene ribbons, $W_{\rm g}$ is the width of the graphene ribbons, $D_{\rm g}$ is the thickness of the graphene ribbons, $L_{\rm g}$ is the trench width, ${\sigma _{0}}$ is the built-in stress of the graphene ribbons,and $Z$ is the deflection at the center of the graphene ribbons, $Z = {\delta _{Z} }+ {\delta _{Z0}}$, where $\delta _{Z0}$ is the position at which the proof mass is balanced under the action of gravity and can also be calculated by equation (2), and $\delta _{Z}$ is the displacement relative to the initial equilibrium position $\delta _{Z0}$.

\begin{figure}[t]
\begin{center}
\includegraphics[width=0.85\textwidth]{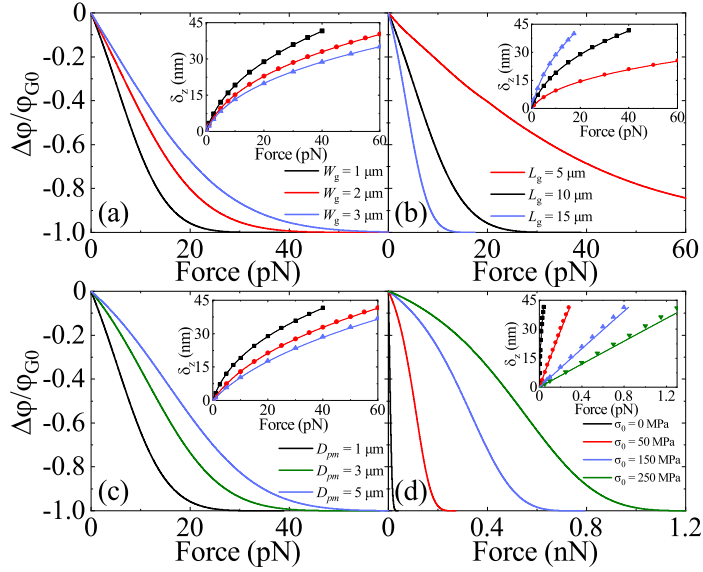}
\end{center}
\caption{Variation of the relative emission intensity of NVC with the applied force for (a) different ribbon widths , (b) different ribbon lengths, (c) different proof masses, and (d) different built-in stresses in graphene ribbons. The insets show the force-displacement curves of the corresponding devices, respectively. The solid lines are the results of the analytical solution calculations, and the dots are the results of the finite element simulations. }
\label{fig2}
\end{figure}

To verify the correctness of the analytical solution, the finite element model of the device was established using ANSYS, to simulate the force-displacement curves of proof masses in different devices. As shown in Fig. \ref{fig2}(a)-(d), the simulation results are consistent with both the analytical solution and ANSYS numerical results. We made a comparison with the experimental results, and the calculated results are also in good agreement with the experimental results.

First, the impact of the width of double-layer graphene ribbons on the device performance was investigated. The size of the proof mass in the calculation is 10 $\upmu$m $\times$ 10 $\upmu$m $\times$ 1 $\upmu$m, and the trench width $L_{\rm g}$ = 10 $\upmu$m. The results show that when their proof masses move the same distance, the force required for the device with a narrower double-layer graphene ribbons is smaller.For example, with the same ${\delta_{Z}}$ of 20 nm, when the width of the double-layer graphene ribbons is 1 $\upmu$m and 3 $\upmu$m, the force is 10.71 pN and 20.24 pN, respectively[Fig. \ref{fig2}(a)]. The reason is when Young's modulus is constant, the narrower the double-layer graphene ribbons, the easier stretching the double-layer graphene ribbons, so the smaller the force required for the proof mass to move the same distance. At this time, the relative emission intensity change of NVC $\Delta \varphi/\varphi_{G0}=(\varphi_{G} - \varphi _{G0})/\varphi_{G0}$ changes with the applied force, as shown in Figure. \ref{fig2}(a), where $\varphi_{G0}$ is the emission intensity of NVC when the proof mass is in the initial position.

When the width of the double-layer graphene ribbons is 1 $\upmu$m, the maximum resolution compared to emission intensity change ${S_{max}} = dF{\rm{/}}d(\Delta \varphi/\varphi_{G0})$ is 14.6 pN, which means that relative emission changes by 1\% when the force changes about 0.146 pN. The photodetector can detect the change of per relative emission intensity as small as 1\%, and high-precision single photon counter photodetector can detect the change of per relative emission intensity lower than 0.1\%\cite{Tian:21}, so the force below 0.146 pN can be detected. When the width of graphene ribbons is 2 $\upmu$m and 3 $\upmu$m, ${S_{max}}$ is 21.6 pN and 26.2 pN, respectively.

Next, the influence of trench width $L_{\rm g}$ was explored, as shown in Fig. \ref{fig2}(b). The size of the proof mass is 10 $\upmu$m $\times$ 10 $\upmu$m $\times$ 1 $\upmu$m, the width of the double-layer graphene ribbons is 1 $\upmu$m. When the proof masses move the same distance, the wider the trench width $L_{\rm g}$, the smaller relative deformation and smaller stress generated in the ribbon, and the smaller external force[the insets of Fig. \ref{fig2}(b)]. Applying identical force, the wider the trench width $L_{\rm g}$ is, the more significantly intensity changes[Fig. \ref{fig2}(b)], and the higher the sensitivity to force of the device. When $L_{\rm g}$ is 15 $\upmu$m, the sensitivity to force of the device is greatly increased, the displacement can reach 40 nm under the force of 17.5 pN, and maximum resolution compared to emission intensity change ${S_{max}}$ is 7.6 pN. When a force of 0.1 pN is applied to the sensor, the maximum change in relative emission intensity of the emitter decreases is about 0.85\%. 0.1 pN is about 1\% of the gravity of one living mammalian cell, which makes it possible for this device to detect quality fluctuations caused by basic cellular processes such as Adenosine  Triphosphate synthesis\cite{martial,adeeous}.

In addition, the effect of the size of proof mass on device performance was also investigated. The results show that the larger proof mass produces a greater initial strain and displacement in the double-layer graphene ribbons under gravity, which makes the stiffness of the double-layer graphene ribbons greater. Under the same external force, the larger the size of the proof mass, the smaller the displacement[the insets of Fig. \ref{fig2}(c)], and the change of the relative emission intensity of corresponding NVC is also smaller[Fig. \ref{fig2}(c)]. When $D_{pm}$ = 5 $\upmu$m, $L_{pm}$ and $W_{pm}$ is 10 $\upmu$m, ${S_{max}}$ is 30.9 pN. When $D_{pm}$ = 1 $\upmu$m, ${S_{max}}$ is 14.6 pN, which is consistent with the previous results.

Previous studies have focused on the piezoresistivity effect of graphene, due to that the strain generated in graphene causes a change of its resistance\cite{caoarab,chun2,Smit20p}. Since the resistance of graphene varies little with a small strain, it is generally necessary to fabricate graphene with a built-in stress to provide the initial strain\cite{furain,zhwow,renhkj}. In order to explore the effect of the built-in stress on this performance, the force-displacement curves of the devices with graphene ribbons fabricated with different built-in stresses were investigated when $L_{\rm g}$ = 10 $\upmu$m, $D_{pm}$ = 1 $\upmu$m, and $L_{pm}$ = 1 $\upmu$m [inset of Fig. \ref{fig2}(d)].The presence of built-in stress significantly increases the stiffness of the ribbons, which makes them become more resistant to deformation.The higher the built-in stress, the smaller the displacement of the proof mass under the same force. When the built-in stress is 250 MPa, the force required for the proof mass to move 40 nm is about 30 times larger than that at 0 MPa. At this point, the greater the built-in stress of the ribbons, the smaller the displacement of proof mass and the change of the relative emission intensity of NVC. Therefore, the reduction of the built-in stress can improve the sensitivity and resolution to the force of the device. Since the thermal expansion coefficient of graphene is negative, the built-in stress of graphene ribbons can be increased or decreased by changing the fabrication temperature. In addition, if the proof mass is higher or lower than the peripheral support column in the process of fabricating and transferring of graphene ribbons, that is, the initial length at the trench of graphene ribbons is longer than the trench, the built-in stress of graphene ribbons can be effectively reduced, thus obtaining higher sensitivity.

\subsection{Acceleration sensor}

The acceleration can be equated to inertial force ${F = M_{pm}a_{pm}}$, where $a_{pm}$ is the externally applied acceleration excitation. Thus, the force sensors can be also used to detect acceleration. The displacement of the proof mass of the device driven by acceleration can still be described as a model of the graphene ribbons fixed at both ends. Hence, driven by the external acceleration excitation, the proof mass moves together with the graphene ribbons, and then the distance between the graphene film and NVC is decreased, which causes a change of the emission intensity of NVC due to the near-field interaction between the graphene and NVC\cite{velipro,reanical}.

Specifically, the device can be regarded as a spring-mass-damping system composed of graphene ribbons and proof mass. The graphene ribbons constitute the spring element of the system. The equivalent spring constant of graphene ribbons $K$ is obtained by taking the derivative of displacement according to equation (2)\cite{Fanphe}, $K = 2\frac{{E_{\rm g}W_{\rm g}D_{\rm g}^3}}{{L_{\rm g}^3}} + 3\frac{{E_{\rm g}W_{\rm g}D_{\rm g}}}{{L_{\rm g}^3}}\delta_{Z0}^2 + 2\frac{{{\sigma _0}W_{\rm g}D_{\rm g}}}{{L_{\rm g}}}$. In this way, the proof mass moves along the z-axis perpendicularly to the surface of graphene ribbons, where the mode frequency of the motion can be approximated by:
\begin{equation}
f_z = \frac{1}{{2\pi }}\sqrt {\frac{K}{{M_{pm}}}}
\end{equation}
The characteristic frequency of the system can be obtained by equation (3). At the same time, ANSYS software is used for modal analysis, and the results are in good agreement.

\begin{figure}[!h]
\begin{center}
\includegraphics[width=0.85\textwidth]{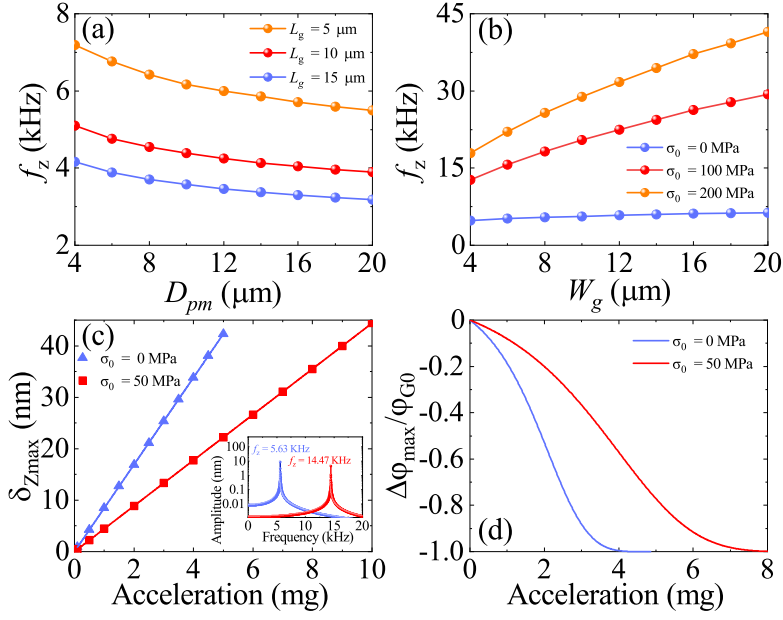}
\end{center}
\caption{(a) Variation of resonant frequency with proof mass thickness for different trench widths $L_{\rm g}$. (b) Variation of resonant frequency with the width of double-layer graphene ribbons under different built-in stresses. (c) Variation of the maximum displacement of the proof mass with acceleration at resonance. The point is the result of finite element simulation, and the solid line is the result of linear fitting. The inset shows the frequency response of the device with different resonant frequencies under 1 mg acceleration excitation. (d) Variation of the maximum relative emission intensity of NVC with acceleration at resonance.}
\label{fig3}
\end{figure}

The effect of the mass of proof mass on the resonant frequency was investigated. With the increase of the thickness of the proof mass $D_{pm}$, the mass of the system increases as well. The oscillator frequency is inversely proportional to the mass of the system, so it decreases as $D_{pm}$ increases. Increasing the length of the double-layer graphene ribbons will reduce the equivalent spring constant significantly, and thereby lead to the decrease of the resonant frequency of the system. For example, when ${L_{\rm g}}$ = 5 $\upmu$m(${L_{\rm g}}$ = 15 $\upmu$m) with the thickness of the proof mass of 10 $\upmu$m, the resonance frequency is 6.17 kHz (3.57 kHz), where the difference between the two is 1.7 times[Fig. \ref{fig3}(a)].

We investigated the effect of the mass of proof mass on the resonant frequency. When the thickness of the proof mass $D_{pm}$ increases, the mass of the system increases, and the oscillator frequency is inversely proportional to the mass of the system, so the oscillator frequency decreases as the thickness of the proof mass $D_{pm}$ increases. When the length of the double-layer graphene ribbons is increased, the equivalent spring constant will be significantly reduced, thereby reducing the system's resonant frequency. For example, when  ${L_{\rm g}}$ = 5 $\upmu$m(${L_{\rm g}}$ = 15 $\upmu$m), the resonance frequency is 6.17 kHz (3.57 kHz) when the thickness of the proof mass is 10 $\upmu$m, and the difference between the two is 1.7 times[Figure. \ref{fig3}(a)].

The spring constant of the ribbons enlarging with the increase of the built-in stress causes the rise of resonance frequency of the system[Fig. \ref{fig3}(b)]. The change of the built-in stress has the most significant effect on the resonant frequency of the system which can be elevated from several kHz to tens of kHz. Thus, increasing the built-in stress of the ribbon is an effective way to detect high frequency vibration signals.

To further show the resonance characteristics of the device, the frequency response of the device was investigated at different built-in stresses at 1 mg acceleration, with the size of proof mass of 20 $\upmu$m $\times$ 20 $\upmu$m $\times$ 17.4 $\upmu$m[inset of Fig. \ref{fig3}(c)]. The displacement response of the proof mass at resonance was obtained[Fig. \ref{fig3}(c)]. When the device resonates, the amplitude of the proof mass is linearly related to the acceleration, whether there is a built-in stress or not. When the built-in stress is 0 MPa, the maximum acceleration resolution compared to  relative emission intensity change $S_{amax} = d{a_{pm}}/d(\Delta \varphi /{\varphi _{G0}})$ is 2.48 mg. At this time, the relative emission intensity change of NVC reaches -1.30\% as the acceleration is 0.1 mg. When the built-in stress is 50 MPa, $S_{amax}$ is 4.72 mg. The relative emission intensity change of NVC reaches -0.67\% as the acceleration is 0.1 mg [Fig. \ref{fig3}(d)].

\begin{figure}[!h]
\flushleft
\begin{center}
\includegraphics[width=0.85\textwidth]{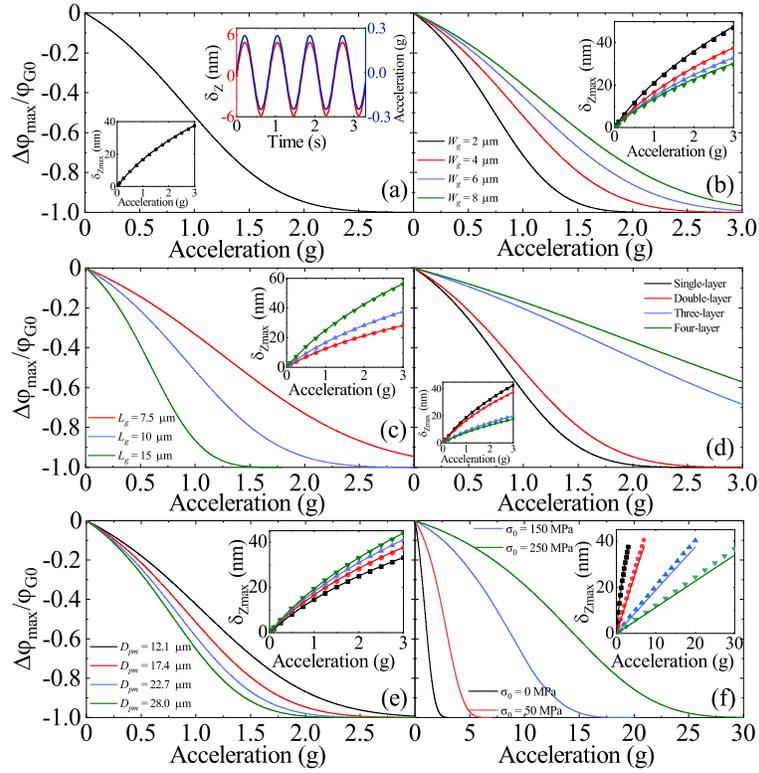}
\end{center}
\caption{(a) Variation of maximum relative emission intensity of NVC with acceleration at 1.2 Hz acceleration. The left inset shows the variation of the maximum displacement with acceleration. The solid line and the point are analytical results and finite element simulation results, respectively. The right inset shows the change of the displacement with time at 0.25 g acceleration. Under (b) different widths of double-layer graphene ribbons, (c) different trench widths Lg, (d) different layers of graphene, and (e) different proof mass. (f) Variation of maximum relative emission intensity of NVC with acceleration  under different built-in stresses. The inset shows  the maximum displacement changes with acceleration.}
\label{fig4}
\end{figure}

To evaluate the non-resonant dynamic characteristics of the device, the maximum displacement change of the proof mass was studied under the sinusoidal acceleration excitation with the frequency of 1.2 Hz[the left inset of Fig. \ref{fig4}(a)]. The results show that the maximum displacement of proof mass also increases with the rise of acceleration, and the numerical simulation results can match well with the analytical calculation results. The displacement of the proof mass is consistent with the dynamic change of the acceleration[the right inset of Fig. \ref{fig4}(a)]. Fig. \ref{fig4}(a) shows the variation of the maximum emission intensity of NVC with acceleration, and $S_{amax}$ reaches 1.57 g.

Next, the effect of the width of double-layer graphene ribbons and trench width ${L_{\rm g}}$ was investigated. When reducing the width of the ribbon or increasing the trench width ${L_{\rm g}}$, the spring constant of the ribbon decreases, and the proof mass becomes easier to move and produces more significant displacement under the same acceleration[the insets of Fig. \ref{fig4}(b) and (c)]. The resonant frequency decreases when the width of the ribbon is reduced or the trench width ${L_{\rm g}}$ is increased. The corresponding change of the maximum relative emission intensity of NVC with acceleration is shown in Fig. \ref{fig4}(b), (c). With the narrower width and the longer length of the ribbon, the device is more sensitive to acceleration. When the width of the ribbon is 2 $\upmu$m (3 $\upmu$m),$S_{amax}$ reaches 1.15 g (0.92 g).

The influence of graphene ribbons of different layers on the performance of the device was studied. Graphene ribbons of different layers have different Young's modulus and thickness, which will significantly impact the performance of the device. Since Young's modulus and thickness of the single- and double-layer graphene ribbons are smaller than those of the three- and four-layer graphene ribbons, the maximum displacement of the proof mass with the former ribbons is larger than that with three- and four-layer graphene ribbons. The maximum displacements of their proof mass are 42.46 nm (17.72 nm) for single graphene (four-layer graphene ribbons) at 1.2 Hz frequency and 3g acceleration[Fig. \ref{fig4}(d)]. Thus, increasing the number of layers of the ribbon, the change of the maximum relative emission intensity of NVC is reduced[Fig. \ref{fig4}(d)]. The larger the size of the proof mass, the greater the force on the graphene ribbons under acceleration, and the greater the resulting proof mass displacement[the inset of Fig. \ref{fig4}(e)]. When the size of the proof mass is 20 $\upmu$m $\times$ 20 $\upmu$m $\times$ 28 $\upmu$m, $S_{amax}$ reaches 1.26 g[Fig. \ref{fig4}(e)].

If there is a built-in stress in the graphene ribbons, the stiffness of the ribbons will be increased, which make it harder for the proof mass to move under acceleration. The built-in stress will increase the resonant frequency, and detuning of drive frequency and resonance frequency increases, which will lead to a decrease of amplitude and a smaller change of maximum relative emission intensity[Fig. \ref{fig4}(f)]. When the built-in stress of the ribbons is 0 MPa, $S_{amax}$ can reach 1.57 g. When the built-in stress of the ribbons is 250 MPa and displacement of the proof mass is 40 nm, the acceleration at the maximum increases about 10 times compared to that without a built-in stress. $S_{amax}$ is 17.2 g If the built-in stress is present, there will be a slight difference (about 8\%) between the results of analytical calculation and the numerical simulation, due to the geometrical nonlinearity of the system structure.

The performance of the designed device was compared with that of existing force and acceleration sensors (as shown in Table\ref{label1} and Table \ref{label2}). The sensitivity to acceleration and the sensitivity to force can be increased by two and seven orders of magnitude, respectively. At the same time, the device has a very small size. But more importantly, after introducing the graphene-induced non-radiative transition into the sensor, the change of the working mode makes the sensor have some special features, such as no need to connect the circuit or power supply, the realization of remote detection and energy charging by light irradiation, thus expanding the application range of the sensor.

\begin{threeparttable}[!hbt]
\scriptsize
\centering
 \caption{The main characteristics of the designed force sensor are compared with those of the force sensor reported in recent years.}
    \begin{tabular}{p{3cm}<{\centering}p{2.5cm}p{2cm}p{2cm}p{2cm}}
    \toprule\rule{0pt}{6pt}
\textbf{Characteristics} & \textbf{Present work} & \textbf{Tianliang}\cite{lirid} & \textbf{Juntian}\cite{qung3d} & \textbf{Cheng}\cite{chedt} \\\midrule
Sensitivity              & 7.01 \%/pN            & 345.2 pm/N           & 525.15 V/N          & 0.27 mV/$\mu$N        \\
Resolution               & $< 0.1$pN   & 2.9 mN               & 4.3 $\upmu$N              & 32 $\upmu$N             \\
Maximum operating range/Resolution  & 1200
& 414            & 581          & 85       \\
Sensing mechanism        & Optical               & Optical              & Piezoresistive      & Piezoresistive  \\\bottomrule[0.75pt]
    \end{tabular}
    \label{label1}
\end{threeparttable}
\\

Finally, the feasibility and scalability of the experiment were discussed. The structure of the designed device is similar to those of the existing experimental fabrications\cite{Fanphe,Fnded,HAurne,blene}. The difference is that the lower surface of the proof mass is attached to a graphene layer, and the base is integrated with a fluorescent emitter(NVC), which is entirely achievable using modern advanced semiconductor manufacturing technology. In addition, graphene and fluorescence emitter manufacturing technology are matured\cite{pedere,schmgh,dohgen,mohity,liuced,rodntal}. The parameters used in the simulation are obtained from experiment, and the difference between the results of numerical simulation and the analytical is slight. Therefore, the results obtained by numerical calculation are achievable in the experiment.

\begin{threeparttable}[!hbt]
\scriptsize
    \centering
     \caption{Comparison of main characteristics of the designed ultra-small NEMS accelerometer and the optical accelerometers reported in recent years.}
   \begin{tabular}{p{3cm}<{\centering}cccc}
    \toprule\rule{0pt}{6pt}
       \bf  Characteristics &{\bf  Present work}&{\bf  Trigona}\cite{trign}&{\bf  Sheikhaleh}\cite{shecro}&{\bf  Soltanian}\cite{soovel} \\\bottomrule[0.75pt]
Dimensions of proof mass ($\upmu$m)&\multirow{2}{*}{$20 \times 20  \times  16.5$}&\multirow{2}{*}{$1900  \times  1900  \times  460$}&\multirow{2}{*}{$250  \times  2500  \times  410$}&\multirow{2}{*}{$11  \times  5  \times  0.8$}\\
Die area occupied by
Functional device parts ($\upmu$m$^2$)& \multirow{2}{*}{  $< 4.8\times 10^3$}&\multirow{2}{*}{$> 1.9  \times  10^7$}&\multirow{2}{*}{$  > 1.1  \times  10^5$}& \multirow{2}{*}{$ > 5.1  \times  10^3$}\\
Optical sensitivity ($|\Delta \varphi/\Delta a|$)&\multirow{2}{*}{41.65 \%/g}&\multirow{2}{*}{0.68 \%/g}&\multirow{2}{*}{0.32 \%/g}&\multirow{2}{*}{0.33 \%/g}\\
Resonance frequency (KHz)&\multirow{2}{*}{3.42}&\multirow{2}{*}{1.97}&\multirow{2}{*}{1.44}&\multirow{2}{*}{17.69}\\\bottomrule[0.75pt]
    \end{tabular}
    \label{label2}
\end{threeparttable}

 \section{Conclusions}
The MEMS force and acceleration sensor based on graphene-induced non-radiative transition was investigated. By utilizing the high sensitivity to distance of the non-radiative transition induced by graphene, a highly sensitive force and acceleration sensor was obtained. The maximum resolution  compared to relative emission intensity change of the external force can reach 7.6 pN. The maximum resolution to acceleration compared to relative emission intensity change can reach 2.48 mg and 0.92 g in the case of resonance and non-resonance respectively. Therefore, this sensor can achieve the detection of a force less than 0.1 pN or an acceleration of 0.1 mg. Increasing the length of the graphene ribbons or reducing the width of the graphene ribbons can further lower its detection threshold for force or acceleration. Reducing the mass of proof mass can improve the sensitivity to force of this structure. Increasing the mass of proof mass can promote its sensitivity to acceleration. This structure, only tens of microns in size, is recharged by light irradiation without connecting a circuit or a power supply, and can realize non-contact remote detection, which makes it can be integrated into the organism. This MEMS force and acceleration sensor will have important application prospects in micro-intelligent devices, wearable devices, biomedical systems, etc.

\section*{Author Contributions}
Guanghui  Li wrote the program, performed the numerical calculation and wrote the paper, F. Liu reviewed the simulation and analyzed the result, S. Yang analyzed the result,  J. Liu, W. Li and Z. Wu supervised the project, designed the structural model, analyzed the results, and wrote the paper.

\section*{Conflicts of interest}
The authors declare that they have no known competing financial
interests or personal relationships that could have appeared to influence
the work reported in this paper.

\section*{Acknowledgments}
This work was supported by National Natural Science Foundation of China (NSFC) (Grants No. 62174040, No. 12174423), the 13th batch of outstanding young scientific and Technological Talents Project in Guizhou Province [2021]5618, the Science and Technology Talent Support Project of the Department of Education in Guizhou Province (Grant No. KY[2018]045),  Guizhou Provincial Science and Technology Projects(Grant No. [2020]1Y026), and the Natural Science Research Project of Department of Education of Guizhou Province (Grant no. QJJ2022015).




\end{sloppypar}
\end{document}